\PassOptionsToPackage{unicode}{hyperref}
\PassOptionsToPackage{hyphens}{url}
\PassOptionsToPackage{dvipsnames,svgnames,x11names}{xcolor}
\documentclass[
  12pt]{article}
\usepackage{algorithm}
\usepackage{algorithmic}
\usepackage{mathtools}

\newcommand{\mysize}{r}
\newcommand{\coneDist}{d_{C}}
\newcommand{\manDist}{\rho}
\usepackage{amsmath,amssymb}
\usepackage{iftex}
\ifPDFTeX
  \usepackage[T1]{fontenc}
  \usepackage[utf8]{inputenc}
  \usepackage{textcomp} 
\else 
  \usepackage{unicode-math}
  \defaultfontfeatures{Scale=MatchLowercase}
  \defaultfontfeatures[\rmfamily]{Ligatures=TeX,Scale=1}
\fi
\usepackage{lmodern}
\ifPDFTeX\else  
\fi
\IfFileExists{upquote.sty}{\usepackage{upquote}}{}
\IfFileExists{microtype.sty}{
  \usepackage[]{microtype}
  \UseMicrotypeSet[protrusion]{basicmath} 
}{}
\makeatletter
\@ifundefined{KOMAClassName}{
  \IfFileExists{parskip.sty}{%
    \usepackage{parskip}
  }{
    \setlength{\parindent}{0pt}
    \setlength{\parskip}{6pt plus 2pt minus 1pt}}
}{
  \KOMAoptions{parskip=half}}
\makeatother
\usepackage{xcolor}
\makeatletter
\ifx\paragraph\undefined\else
  \let\oldparagraph\paragraph
  \renewcommand{\paragraph}{
    \@ifstar
      \xxxParagraphStar
      \xxxParagraphNoStar
  }
  \newcommand{\xxxParagraphStar}[1]{\oldparagraph*{#1}\mbox{}}
  \newcommand{\xxxParagraphNoStar}[1]{\oldparagraph{#1}\mbox{}}
\fi
\ifx\subparagraph\undefined\else
  \let\oldsubparagraph\subparagraph
  \renewcommand{\subparagraph}{
    \@ifstar
      \xxxSubParagraphStar
      \xxxSubParagraphNoStar
  }
  \newcommand{\xxxSubParagraphStar}[1]{\oldsubparagraph*{#1}\mbox{}}
  \newcommand{\xxxSubParagraphNoStar}[1]{\oldsubparagraph{#1}\mbox{}}
\fi
\makeatother

\usepackage{longtable,booktabs,array}
\usepackage{calc} 
\usepackage{etoolbox}
\makeatletter
\patchcmd\longtable{\par}{\if@noskipsec\mbox{}\fi\par}{}{}
\makeatother
\IfFileExists{footnotehyper.sty}{\usepackage{footnotehyper}}{\usepackage{footnote}}
\makesavenoteenv{longtable}
\usepackage{graphicx}
\makeatletter
\def\maxwidth{\ifdim\Gin@nat@width>\linewidth\linewidth\else\Gin@nat@width\fi}
\def\maxheight{\ifdim\Gin@nat@height>\textheight\textheight\else\Gin@nat@height\fi}
\makeatother
\setkeys{Gin}{width=\maxwidth,height=\maxheight,keepaspectratio}
\makeatletter
\def\fps@figure{htbp}
\makeatother

\makeatletter
\@ifpackageloaded{caption}{}{\usepackage{caption}}
\AtBeginDocument{%
\ifdefined\contentsname
  \renewcommand*\contentsname{Table of contents}
\else
  \newcommand\contentsname{Table of contents}
\fi
\ifdefined\listfigurename
  \renewcommand*\listfigurename{List of Figures}
\else
  \newcommand\listfigurename{List of Figures}
\fi
\ifdefined\listtablename
  \renewcommand*\listtablename{List of Tables}
\else
  \newcommand\listtablename{List of Tables}
\fi
\ifdefined\figurename
  \renewcommand*\figurename{Figure}
\else
  \newcommand\figurename{Figure}
\fi
\ifdefined\tablename
  \renewcommand*\tablename{Table}
\else
  \newcommand\tablename{Table}
\fi
}
\@ifpackageloaded{float}{}{\usepackage{float}}
\floatstyle{ruled}
\@ifundefined{c@chapter}{\newfloat{codelisting}{h}{lop}}{\newfloat{codelisting}{h}{lop}[chapter]}
\floatname{codelisting}{Listing}

\makeatother
\makeatletter
\@ifpackageloaded{caption}{}{\usepackage{caption}}
\@ifpackageloaded{subcaption}{}{\usepackage{subcaption}}
\makeatother

\ifLuaTeX
  \usepackage{selnolig}  
\fi
\usepackage[]{natbib}
\usepackage{bookmark}

\IfFileExists{xurl.sty}{\usepackage{xurl}}{} 
\hypersetup{
  pdftitle={Title},
  pdfauthor={Author 1; Author 2},
  pdfkeywords={3 to 6 keywords, that do not appear in the title},
  colorlinks=true,
  linkcolor={blue},
  filecolor={Maroon},
  citecolor={Blue},
  urlcolor={Blue},
  pdfcreator={LaTeX via pandoc}}

\newcommand{\anon}{1}

\begin{document}

\date{}

\def\spacingset#1{\renewcommand{\baselinestretch}%
{#1}\small\normalsize} \spacingset{1}


\if1\anon
{
  \title{\bf Principal Nested Cones}
  \author{Yanyan Zhan\\
    Department of Statistics, University of South Carolina\\
    Ian L.~Dryden\\
    School of Mathematical Sciences, University of Nottingham\\
    and \\
    Yuexuan Wu\\
    Department of Statistics, University of South Carolina\\}
  \maketitle
} \fi

\if0\anon
{
  \bigskip
  \bigskip
  \bigskip
  \begin{center}
    {\LARGE\bf Principal Nested Cones}
\end{center}
  \medskip
} \fi

\bigskip
\begin{abstract}
In many applications, the data lie on a type of cone, where there is a distinction between an overall scale variable and the remaining scale-free structure. For example, the joint size and shape of objects are points on a cone, where size represents scale, and shape is the scale-free structure. Dimension reduction is central in such applications, as shape data are often high-dimensional. Interactions between shape and size are widespread and of significant interest in real-world applications. However, most existing methods either lack a single notion of size or focus solely on shape, effectively removing size information. We propose Principal Nested Cones (PNC), a nonlinear dimension reduction framework that preserves both shape and size. PNC represents data through a sequence of nested hypercones and progressively projects observations onto lower-dimensional cone spaces. The resulting PNC scores provide low-dimensional representations that jointly capture size–shape variation in an interpretable manner. To enable scalable computation in ultra-high-dimensional settings, we develop a fast approximation combining PCA-based transformation with standard PNC. Simulation studies and real data applications demonstrate that PNC captures nonlinear size--shape structure, improves representation and reconstruction, and yields interpretable insights across morphometric, developmental, and molecular datasets. 
\end{abstract}

\noindent%
\vfill

\newpage
\spacingset{1.5} 

\section{Introduction}

{
In many statistical applications, it is common to encounter data that lie on a type of cone, where there is a distinction between an overall scale variable and 
the remaining scale-free structure. For example, in biology, it is of interest to examine the relationship between size 
(a scale variable) and shape (geometrical properties which are scale-free). Another example is a covariance matrix $\Sigma$ which can be represented as the total variation ${\rm tr}({\Sigma})$ and a scale-free trace 1 positive definite matrix  $\Sigma/{\rm tr}({\Sigma})$. The distinction between scale and scale-free structure leads to the representation of the data on a cone of a metric space, which 
is the product of a positive real line (the scale) and an underlying metric space $M$. The use of the cone representation can be helpful for examining the joint relationship between the scale variable and data on the underlying metric space $M$. 
}

Among such settings, object and shape data arise in a wide range of applications, including biology, medical imaging, computer vision, and molecular analysis 
\citep{dryden2016statistical,Marrdryd22}. These data are commonly represented by landmark coordinates or derived measurements such as lengths, widths, and angles, and can therefore be high-dimensional. In addition, complex interactions between shape and size are common. Dimension reduction and joint size-and-shape analysis are therefore fundamental challenges in statistical modeling.

Dimension reduction can be achieved through feature selection or feature projection. While feature selection is sometimes practical for measurement shape data, projection-based methods are more commonly used in general settings. Among projection-based methods, principal component analysis (PCA) is the standard approach \citep{hotelling1933analysis, jolliffe02,abdi2010principal}. Since shape space is inherently non-Euclidean, PCA is typically performed in the tangent space of shape or size-and-shape space, a procedure referred to as \textit{shape PCA} \citep[Section 7.7]{dryden2016statistical}. 
Another example is the elastic shape analysis framework, which represents shapes using invariant representations such as square-root velocity functions, and performs dimension reduction via PCA in the associated tangent space \citep{srivastava2016functional, wu2024shape}. 

Principal Nested Spheres (PNS; \citealp{Jungetal12}) provides an alternative by performing a sequence of projections onto nested spherical submanifolds rather than relying on tangent space approximations. However, PNS removes scale by normalizing observations to unit size, thereby discarding information about size.

In many applications, shape and size are intrinsically linked, and their interaction is widespread in nature. A common approach is to separate size and shape, and then analyze their relationship using statistical tools. For example, allometry studies the dependence of shape on size, typically by treating size as a predictor and shape as the response \citep{klingenberg2016size}. 
Another approach, based on \textit{Procrustes form space}, augments the tangent space coordinates with the logarithm of centroid size and applies PCA to the combined representation.
However, such approaches do not constitute a fully joint analysis of size and shape, since size and shape are separated prior to modeling and then combined as independent components. 

A more natural approach is to retain \textit{both shape and size} throughout the analysis, so that no artificial separation is introduced. In settings where overall size differences are believed to dominate the variance, one common exploratory approach is to apply PCA directly to size-and-shape data and interpret the first principal component as representing size and the remaining components as representing shape. However, this interpretation is subjective and lacks a clear geometric justification. More generally, dimension-reduction methods aim to capture the dominant sources of variation in the first few components, while progressively removing less informative variation in subsequent steps. 

Motivated by these limitations, we propose principal nested cones (PNC), a nonlinear dimension-reduction framework that jointly models size and shape.  PNC generalizes PNS from shape space to size-and-shape space. By a sequence of nested cones, size is retained as an explicit geometric component throughout the analysis. It leads to a novel representation for size-and-shape data from a scale--structure decomposition, in which a radial component encodes size and a manifold-valued component encodes shape. Moreover, the approach generalizes naturally to broader Euclidean and non-Euclidean settings.


The remainder of the paper is organized as follows. Section~2 introduces the general cone spaces, examples, and their geometric properties. Section 3 develops the PNC framework in the hypercone setting with a detailed modeling procedure. Section 4 presents simulation studies, followed by real data applications in Section 5. Section 6 concludes with a discussion.

\section{Cone space}
\label{sec:conespace}
We first introduce the notion of cone spaces, which provide a natural geometric framework for size--structure data. 

\subsection{Cone of a metric space}
A cone ${\rm Con}(M)$ of a metric space consists of the product 
of a metric space $(M,d)$ with the non-negative real numbers 
$[0,\infty)$. Each point on the cone can be written as $(x,t)$ where $x \in M$ and $t \ge 0$, and in addition all points 
$(x,0)$ are identified with each other. The cone
is therefore a quotient space $\{ M \times [0,\infty) \}/ \sim$, 
where the apex equivalence relation $\sim$ identifies all
$(x_1,0) \sim (x_2,0)$ as the single apex point.

\textbf{Result 1.} {\it Let $M$ be a metric space with diameter at most $\pi$, then the cone metric $\coneDist$ on ${\rm Con}(M)$ is  
\begin{equation}
\coneDist(p,q) = \sqrt{ t^2 + s^2 -2 t s \cos( \manDist(x,y) ) } \, , 
\label{eq:geodist}
\end{equation}
where $p,q \in {\rm Con}(M)$ and $p = (x,t), q=(y,s)$. 
}

\textbf{Result 2.} {\it  If $\manDist(x,y)$ is an intrinsic distance then $\coneDist$ is an intrinsic distance on the cone. }

Both these results are provided in the textbook of 
\cite[pp91-92]{Buragoetal01}.

\subsection{Cone manifolds}
The cone construction above is not restricted to spheres. More generally, let $M$ be a compact Riemannian manifold with intrinsic geodesic distance $\manDist(\cdot,\cdot)$, which ensures well-defined local projections onto submanifolds. Then the quotient
\(
\mathrm{Con}(M) = M \times \mathbb{R}^+ / \sim
\)
is the cone over $M$ and provides a natural size--structure space.

\textbf{Result 3.} {\it The metric tensor on the cone manifold is a warped product metric}
\begin{equation*} 
ds^2 = dr^2 + r^2 g \, ,
\end{equation*}
{\it where \( r \in \mathbb{R}^+ \) is the radial coordinate, and \( g \) denotes the Riemannian metric tensor on \( M \).}

Under this general formulation, we can construct a nested sequence of submanifolds 
\[M = M_d \supset M_{d-1} \supset \cdots \supset M_1,\] which induces a nested sequence of cone submanifolds
\[
\mathrm{Con}(M) = \mathrm{Con}(M_d) \supset \mathrm{Con}(M_{d-1}) \supset \cdots \supset \mathrm{Con}(M_1).
\]
To estimate such a sequence of submanifolds, at each stage $k$, we can minimize the corresponding empirical objective function
\begin{equation}
\label{eq:genericobj}
    Q_{n,k} = \sum_{i=1}^{n} r_i^2 \, \manDist \left(x_i, \pi_k(x_i) \right)^2,
\end{equation}
where \((x_i,r_i)\in \mathrm{Con}(M)\), $r_i \ge 0$ is the size of observation $x_i$, and $\pi_k$ is the intrinsic projection onto the lower-dimensional submanifold $M_{k-1}$. Thus, in the general setting, we minimize a size-weighted sum of squared intrinsic distances on the manifold $M$, where size corresponds to the radial coordinate. The detailed methodology developed later specializes this general formulation to the spherical case, where the cone admits an explicit Euclidean realization via hypercones and the criterion reduces to the sum of squared hypercone residuals.






\subsection{Hypercone}
A particularly important case of such a cone space is a \textbf{hypercone} where the metric space is 
a sphere of radius $\sin \alpha$ in $\mathbb{R}^d$, 
where $\alpha \in [0,\frac{\pi}{2}]$. 
Hence $M = S^{d-1}( \sin\alpha )$ is the metric space and 
the distance in the metric space
is $\sin \alpha \cdot \Theta$ where $\Theta$ is the usual great circle distance on the unit sphere.

\begin{figure}[ht!]
\begin{center}
    \includegraphics[width=1\linewidth]{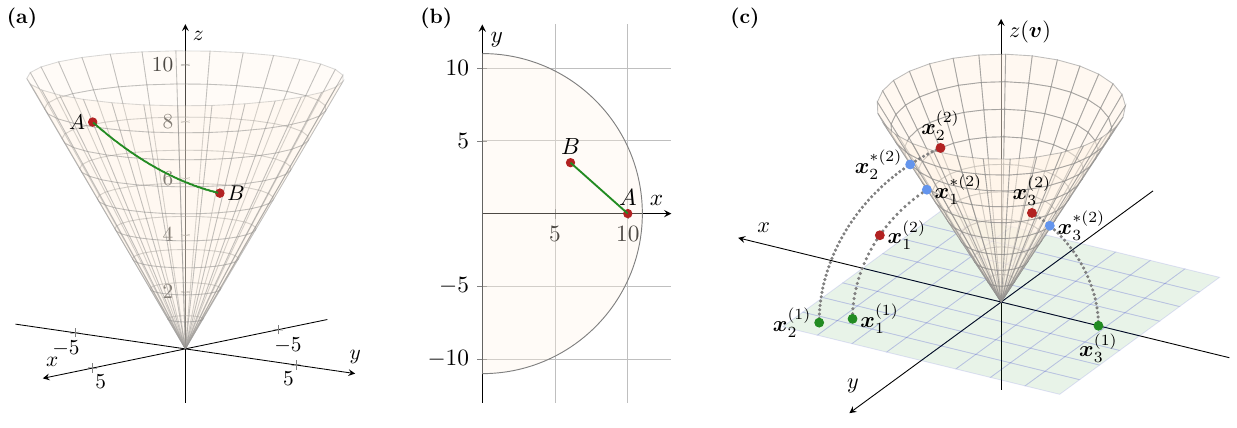}
\end{center}
\caption{(a) An example of a 3D cone with 
opening angle $\alpha = \pi/6$. A minimal 
geodesic path is drawn between points at sizes
$r=7$ and $r=10$, separated by $\Theta = \pi/3$ on the unit sphere. The length of
the geodesic is 
$d_C = \sqrt{ 10^2 + 7^2 - 2 \cdot 10 \cdot 7 \cdot \cos( \sin \pi/6 \cdot \pi/3) } =  5.2684$. 
(b) The flattened cone, where the geodesic becomes a straight line of the same length as the geodesic on the cone.
(c) Illustration of projection and mapping functions in 3D. Red points: input $\boldsymbol{x}^{(2)}_j$; Blue points: projected points $\boldsymbol{x}^{*(2)}_j = \operatorname{P}(\boldsymbol{x}^{(2)}_j, \boldsymbol{v}, \alpha)$; Green points: mapped points to 2D: $ \boldsymbol{x}^{(1)}_j = \operatorname{M}(\boldsymbol{x}, \boldsymbol{v}), \; \; j=1,2,3$.}
\label{fig:3dexample}
\end{figure}

In more detail, a hypercone in \( \mathbb{R}^{d+1} \) with apex at the origin and axis along a unit vector \( \boldsymbol{v} \in \mathbb{R}^{d+1} \) is defined as the set of vectors $\boldsymbol{x}$ that form a constant angle \(\alpha\) with \(\boldsymbol{v}\), and consists of two nappes corresponding to $\cos \alpha \ge 0$ and
$\cos \alpha \le 0$ respectively. We focus on the upper half hypercone with one nappe, given by
\begin{equation}
	\label{eq:hypercone}
\mathcal{C}_d(\boldsymbol{v}, \alpha) = \left\{ \boldsymbol{x} \in \mathbb{R}^{d+1} \setminus \{\boldsymbol{0}\} \;:\; \frac{\boldsymbol{x}^T \boldsymbol{v}}{\mysize} = \cos(\alpha) \ge 0 \right\},
\end{equation}
where \( \alpha \in [0, \frac{\pi}{2}] \) is the fixed opening angle, \(\mysize = \|\boldsymbol{x}\| = \sqrt{\boldsymbol{x}^T \boldsymbol{x}} \) is the distance to the apex of the hypercone, i.e., the slant length. Hypercones are differentiable regular 
manifolds except at the apex, which is a singular point.  An example coordinate system for a \(d\)-dimensional hypercone with an axis $\boldsymbol{v}$ along the \(x_{d+1}\)-axis is
\begin{align}
x_1 &= \mysize \sin(\alpha) \cos(\theta_1) \nonumber \\
x_k &= \mysize \sin(\alpha)  \left(\prod_{i = 1}^{k-1} \sin(\theta_i) \right)  \cos(\theta_k) \text{ for } k \in \{2, \cdots, d-1\} \nonumber \\
x_{d} &= \mysize \sin(\alpha) \prod_{i = 1}^{d-1} \sin(\theta_i) \nonumber \\
x_{d+1} &= \mysize \cos(\alpha) 
\label{eq:coords}
\end{align}
where \(\theta_1, \dots, \theta_{d-2} \in [0, \pi]\), \(\theta_{d-1} \in [0, 2\pi]\), and \(\alpha \in [0, \pi/2]\).
The hypercone is easy to visualize when $d=2$, which is the familiar cone in 3D, as shown in Figure~\ref{fig:3dexample}~(a). In that case, we have the z-axis of the cone given by $z = \mysize \cos \alpha$, 
and the $x$-$y$ coordinates lie on the circle $S^1$ with radius $\mysize \sin\alpha$. 
The same idea generalizes to higher dimensions, 
where the first $d$ coordinates lie on a hypersphere of radius $\mysize \sin \alpha$ at a height on the $(d+1)$th coordinate of $\mysize \cos \alpha$.

\textbf{Hypercone geodesics.}
\label{sec:conegeo}
The minimal geodesic path minimizes the 
intrinsic distance between two points in the hypercone. Using (\ref{eq:geodist}), we have
the following expression for the geodesic distance:

\textbf{Result 4:} 
{\it Given two points $\mathbf{x}_1, \mathbf{x}_2$ in a hypercone, the intrinsic distance between the points is the length of the minimal geodesic path
which is given by
\begin{equation}
\coneDist( \mathbf{x}_1, \mathbf{x}_2 ) = 
\sqrt{ \| \mathbf{x}_1 \|^2 + 
 \| \mathbf{x}_2 \|^2 - 2 \| \mathbf{x}_1 \| 
 \| \mathbf{x}_2 \| \cos \left( \sin (\alpha) \cdot \Theta \right) }  \;,
 \label{eq:intrinsic}
\end{equation}
where $\Theta \in [0,\pi]$ is the shortest great circle distance on the 
sphere between angular components of $\mathbf{x}_1$ and 
$\mathbf{x}_2$ on $\Omega = S^{d-2}$.
}


\textbf{Proof (sketch).} The result is a special case of Result 1. For intuition, we outline a geometric argument based on an isometric flattening of the hypercone here. A full proof is provided in the supplementary material.

The hypercone is equipped with the warped product metric
$ds^2 = dr^2 + r^2 \, \sin^2 (\alpha) \,d\Omega^2,$
where $d\Omega$ is the standard metric on the unit sphere. By rescaling the angular component via \( \theta \mapsto \sin(\alpha)\,\theta \), the hypercone can be mapped isometrically to a Euclidean hypersector. Under this flattening transformation, geodesics on the hypercone correspond to straight lines in the flattened Euclidean space. Consequently, the geodesic distance between two points is given by the usual Euclidean Law of Cosines, with the angular distance replaced by the rescaled spherical distance $\sin(\alpha)\, \Theta$. \hfill $\Box$

To compute geodesic paths on the cone, we first map the space isometrically to a flattened hypersector, where geodesics correspond to straight lines, and then transform back to the intrinsic coordinates~\eqref{eq:coords}. 
Figure~\ref{fig:3dexample}~(a, b) shows an example of a hypercone in $\mathbb{R}^3$ and a geodesic path between two points, both in the original hypercone in 3D and in the flattened cone, which is a sector.  



%
\subsection{Kendall's size-and-shape cone}

An important non-spherical example is \textbf{Kendall’s size-and-shape space} \citep{Kendall89}. Let $\Sigma_m^k$ denote Kendall’s shape space \citep{kendall1984shape} of $k$ landmarks in $\mathbb{R}^m$, which is obtained as the quotient of the preshape sphere under the rotation group. The intrinsic distance on $\Sigma_m^k$ is the Procrustes geodesic distance. The size-and-shape cone is then defined as the quotient $\mathrm{Con}(\Sigma_m^k) = \Sigma_m^k \times \mathbb{R}^+ / \sim$, where all points $(x,0)$ are identified as a single apex and the radial coordinate corresponds to the centroid size. 

It is convenient to construct the nested cones in this size-and-shape space using an ambient preshape representation that retains scale. Let $\mathbf{X} \in \mathbb{R}^{m \times k}$ denote a mean-centered landmark configuration with quotiented rotation but not normalized to unit size (size kept). The centroid size is
\(
r = \|\mathbf{X}\|_F = \sqrt{ {\rm tr}( \mathbf{X}^T \mathbf{X}) } ,
\)
where $\|\cdot\|_F$ denotes the Frobenius norm.
The corresponding preshape is then 
\[
\mathbf{Z} = \frac{\mathbf{X}}{\| \mathbf{X} \|_F}.
\]
That is, each configuration can be represented by the pair $(\mathbf{Z}, r)$, where $\mathbf{Z}$ encodes shape (modulo rotation) and $r$ encodes size. Nested size-and-shape cones are of interest in molecule dimension reduction as seen in Section \ref{sec:moldata}.



\subsection{Symmetric positive definite cone}
In the same spirit, the scale--structure decomposition extends naturally to covariance matrices. For a \textbf{symmetric positive definite (SPD)} covariance matrix $\Sigma \succ 0$, we define
\[
\Sigma = s \, \tilde{\Sigma}, \quad s = \mathrm{tr}(\Sigma), \quad \text{and} \quad \mathrm{tr}(\tilde{\Sigma}) = 1.
\]
Here $s$ represents total variance (size) and $\tilde{\Sigma}$ represents the normalized covariance structure.

The resulting space 
$
\mathcal{S}_{++}^q (1) = \{ \tilde{\Sigma} \succ 0 : \mathrm{tr}(\tilde{\Sigma}) = 1\}
$ 
forms a smooth manifold under intrinsic Riemannian metrics, such as the affine-invariant \citep{pennec2006riemannian} or log-Euclidean metrics \citep{arsigny2006log}. The cone is then defined as
\[
\mathrm{Con}(\mathcal{S}_{++}^q (1)) = \mathcal{S}_{++}^q (1) \times \mathbb{R}^+ / \sim,
\]
which keeps both the scale and the covariance geometry. Thus, nested cones in this setting correspond to restricting structural variation $\tilde{\Sigma}$ to lower-dimensional submanifolds that express reduced structural variability.

\textbf{Remark.}
The examples above show that the nested cone construction provides a unified geometric framework for size--structure data. In particular, the spherical cone admits an explicit Euclidean realization through hypercones, whereas the Kendall's size-and-shape space and the SPD matrices are non-Euclidean examples of the same general construction. It is also worth noting that the radial coordinate in $\mathbb{R}^+ $ need not correspond strictly to geometric size, but may represent a general positive scale parameter, such as duration or intensity. For example, in survival analysis, observations can be represented as $(x_i, t_i)$, where $t_i > 0$ denotes time-to-event (a duration scale).
The detailed methodology developed in the next section focuses on the hypercone case, where the geometry is most explicit. Nevertheless, the same underlying framework extends naturally to these more general settings.




\section{Principal nested cones}
\label{sec:PNC}
\subsection{General framework}
Building on the cone space framework introduced in Section~\ref{sec:conespace}, we now specialize to the spherical case, where the cone admits explicit Euclidean realization through \textit{hypercones}.

PNC constructs a sequence of nested hypercones from higher- to lower-dimensional spaces to perform dimension reduction through nonlinear transformation. It proceeds via a backward, progressive scheme that removes less informative variation while preserving size information. This contrasts with the usual forward derivation of PCA, which identifies dominant directions of variation (although PCA 
can also equivalently be viewed with a backward derivation,  
\cite{Marronetal10}). 
PNC reveals the dominant structure by successively removing less significant components.

We define the size (radial coordinate) of an observation $\boldsymbol{x}$ as $ \mysize = \left\lVert \boldsymbol{x} \right\rVert = \sqrt{\boldsymbol{x}^T \boldsymbol{x}}$. In the hypercone model, this equals the Euclidean distance from the origin (the apex) to the observation. For mean-centered landmark configurations, it coincides with the standard centroid size \citep{kendall1984shape}.

\begin{figure}[ht!]
	\centering
	\includegraphics[width=1\linewidth]{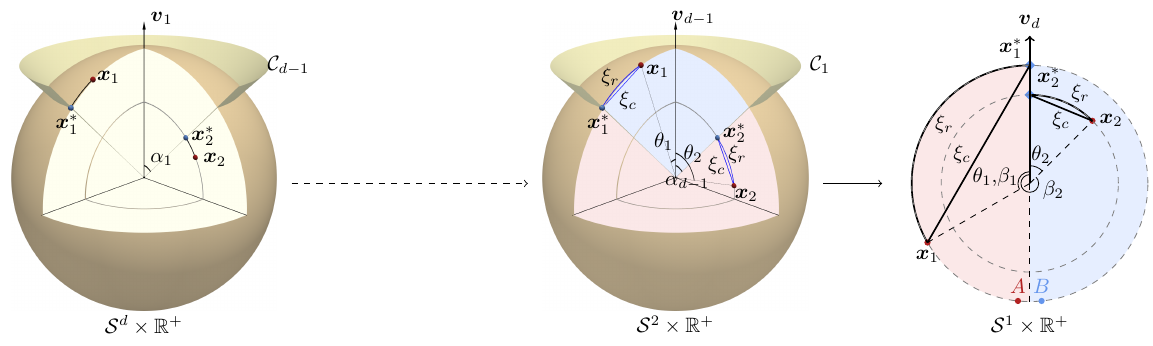}
	\caption{Illustration of PNC modeling. Vectors \( \boldsymbol{x}_i\) denote observations and \( \boldsymbol{x}^*_i\) their projections onto hypercone surfaces.
    From Steps $1$ to $d-1$, observations inside the cones have negative scores and residuals, while those outside are positive. At Step $d$, observations clockwise from the cone axis are negative, and those counterclockwise are positive.
    }
	\label{fig:PNC}
\end{figure}

The framework of PNC is illustrated in Figure~\ref{fig:PNC}, which follows the general backward decomposition idea of PNS \citep{Jungetal12}.
Consider a random sample of $n$ data vectors  
in $\mathbb{R}^{d+1}$. The data matrix \( \boldsymbol{X} \) has dimensions \( (d+1) \times n \), where each column represents an observation. Dimension reduction proceeds over $d$ stages to reduce the data from \( (d+1) \) to \( 2 \) dimensions. To track the data across stages, we denote \( \boldsymbol{X}^{(d+1-k)} \) the input at Step \( k \), which has dimensions \( (d+2-k) \times n \). In particular, \( \boldsymbol{X}^{(d)} = \boldsymbol{X} \). While the dimensionality of \( \boldsymbol{X}^{(d+1-k)} \) decreases with \(k\), the size of each observation remains unchanged throughout. 

At Step~1, $ \mathcal{S}^d \times \mathbb{R}^+ / \sim $ is reduced to $ \mathcal{S}^{d-1} \times \mathbb{R}^+ / \sim $ via the hypercone \( \mathcal{C}_{d-1} \). The resulting matrix \( \boldsymbol{X}^{(d-1)} \) becomes the input for Step~2, and the procedure proceeds iteratively.
In general, at Step \( k \), the data matrix \( \boldsymbol{X}^{(d+1-k)} \) lies in $ \mathcal{S}^{d+1-k} \times \mathbb{R}^+ / \sim$ for \( k=1,\ldots,d - 1\), and a hypercone \( \mathcal{C}_{d-k} \) is fitted. The observations are then mapped to $ \mathcal{S}^{d-k} \times \mathbb{R}^+ / \sim $, yielding \( \boldsymbol{X}^{(d-k)} \).
The final step differs slightly: the 2D cone angle is set to 0, reducing the sector to a one-dimensional direction (i.e., a vector).

\subsection{Model construction}
The core of PNC modeling is to identify a sequence of hypercones that serve as intermediate structures linking the current space to a lower-dimensional space. We first introduce the key components in logical order, and then summarize the full procedure.

\textbf{Nested cones.}
At Step \( k \), a nested cone is a sub-manifold of \(\ \mathcal{S}^{d+1-k} \times \mathbb{R}^+ / \sim \) that captures the dominant pattern in the data. Each cone is determined by an axis direction \( \boldsymbol{v}_k\) and an opening angle \( \alpha_k\). 
Here $\boldsymbol{v}_k \in \mathbb{R}^{d+2-k}$ is a unit vector, $\alpha_k \in [0,\pi/2]$ is a scalar angle.
The corresponding hypercone is denoted by \(\mathcal{C}_{d - k} = \mathcal{C}(\boldsymbol{v}_k, \alpha_k)\).


\textbf{Projection function.}
The projection function \( \operatorname{P}(\boldsymbol{x}, \boldsymbol{v}, \alpha)\) maps \( \boldsymbol{x} \) onto the cone \(\mathcal{C}(\boldsymbol{v}, \alpha)\), as illustrated in Figure~\ref{fig:3dexample}~(c). The projected point \( \boldsymbol{x}^* \) remains in the same ambient space and preserves the size $\mysize$.
The function \( \operatorname{P}( \cdot) \) is defined using spherical linear interpolation (SLERP; \citealp{shoemake1985animating}) as 
\begin{equation}
	\label{eq:slerp}
	\boldsymbol{x}^*  = \operatorname{P}(\boldsymbol{x}, \boldsymbol{v}, \alpha) 
    = \frac{\sin(\alpha)\, \frac{\boldsymbol{x}}{\mysize} + \sin(\theta - \alpha)\, \boldsymbol{v}}{\sin(\theta)}\, \mysize ,
\end{equation}
where  \( \theta =
\arccos\!\left( \boldsymbol{x}^T \boldsymbol{v} / {r}
\right) \).


The projection \eqref{eq:slerp} is well-defined for \(\theta \in (0, \pi) \). The boundary cases where \(\theta = 0 \) or \(\theta = \pi \) correspond to observations that are exactly aligned or anti-aligned with $\boldsymbol{v}$. Such configurations form a set of measure zero under any absolutely continuous distribution on the sphere and are therefore negligible in practice.


\textbf{Residuals.}
Residuals quantify the deviation of observations $\boldsymbol{x}$ from the fitted cone, defined as signed distances between \(\boldsymbol{x}\) and its projection \(\boldsymbol{x}^*\). Intuitively, they measure the information that is lost during projection. 
Two forms of residuals $\xi$ are considered: intrinsic Riemannian distance (great circle arcs $\xi_r$  in Figure~\ref{fig:PNC}) and chordal extrinsic distance ($\xi_c$ in Figure~\ref{fig:PNC}). From Step 1 to Step \( (d - 1) \),
\begin{equation*}
	\label{eq:resid}
	\xi_r = (\theta - \alpha)\, \mysize  \quad \text{and} \quad
		\xi_c  = 2 \sin \left(\frac{\theta - \alpha}{2}\right) \mysize \, \,  ,
\end{equation*}
where \(\theta\) is the angle between the observation \(\boldsymbol{x}\) and the cone axis $\boldsymbol{v}$. Observations inside the cone have negative residuals, while those outside have positive residuals.
At the final stage ($\alpha_d = 0$), residuals reduce to
\begin{equation*}
	\label{eq:resid-d}
	\xi_r = \theta\, \mysize  \quad \text{and} \quad
	\xi_c = 2 \sin \left(\frac{\theta}{2}\right)\, \mysize \,  ,
\end{equation*}
where \(\theta\) is the signed angle between the observation \(\boldsymbol{x}\) and the cone axis \(\boldsymbol{v}_d\), computed as
\begin{equation*}
	\theta = \operatorname{atan2}\!\big( \det[\boldsymbol{v}_d,\boldsymbol{x}],\; \boldsymbol{x}^T \boldsymbol{v}_d \big) = {\rm Arg}( \boldsymbol{x}^T \boldsymbol{v}_d + i \det[\boldsymbol{v}_d,\boldsymbol{x}] ) \, .
\end{equation*}
For those at the clockwise side of the 2D cone axis, residuals are negative, while for those at the counterclockwise side, residuals are positive.

The defined residuals incorporate size and are computed between the original points and their projections. 
Compared with \(\xi_r\), the chordal distance \(\xi_c\) reduces weights on observations away from nested cones when computing objective functions, and the reduction is more obvious on those with larger sizes. When the residuals are small, the two distances are similar. Unless stated otherwise, the Riemannian distance \(\xi_r\) will be used for the following analysis.

The residual vector contains residuals for all $n$ observations and is an $n$-element vector, denoted by \(\boldsymbol{\xi}\). Subscripts will be used when necessary to stress which steps they are from. Specifically, \(\boldsymbol{\xi}_{d + 1 - k}\) is a \(n \times 1\) vector from Step \(k\). The collection of residuals from all \(d\) steps, \( [\boldsymbol{\xi}_d, \cdots,  \boldsymbol{\xi}_1 ]\) , is denoted as \(\boldsymbol{\Xi}\), which forms an \(n \times d\) matrix.

\textbf{Mapping function.}
The mapping function \(  \operatorname{M}(\boldsymbol{x}, \boldsymbol{v}) \) maps \(\boldsymbol{x}\) from \(\ \mathcal{S}^{d+1-k} \times \mathbb{R}^+ / \sim \) to \(\ \mathcal{S}^{d-k} \times \mathbb{R}^+ / \sim \), as illustrated in Figure~\ref{fig:3dexample}~(c). It does not take \(\alpha \) as an input, which means opening angles do not influence the mapping outcome. The function reduces dimensionality by one while preserving size, and consists of two parts: 

\textit{(i) Rotation:} Rotate the coordinate system such tah  $\boldsymbol{v}$ is aligned with a chosen standard direction $\boldsymbol{e}$. The rotated vector is $
	\boldsymbol{x}_{\text{rot}} = \boldsymbol{R} \boldsymbol{x}$,
and \(\boldsymbol{R}\) is Rodrigues' rotation matrix \citep{murray2017mathematical} determined by \(\boldsymbol{v}\) and can be computed by 
\begin{equation*}
	\label{eq:rotation}
	\boldsymbol{R} = \boldsymbol{I} + \sin(\gamma) (\boldsymbol{e}_m \boldsymbol{c}^T - \boldsymbol{c} \boldsymbol{e}_m^T) + \left(\cos(\gamma) - 1 \right) (\boldsymbol{e}_m \boldsymbol{e}_m^T + \boldsymbol{c} \boldsymbol{c}^T),
\end{equation*}
where $\boldsymbol{e}_m = [0,\ldots,0,1]^T$, 
$v_m = \boldsymbol{v}^T \boldsymbol{e}_m$, 
$\gamma = \arccos(v_m)$ 
and 
$c = \frac{v - v_m \boldsymbol{e}_m}{\|v - v_m \boldsymbol{e}_m\|}$.

\textit{(ii) Reduction:} Remove the last component of the rotated vector and rescale the remaining components to preserve magnitude.



\subsection{Low-dimensional representation}
\label{sec:low-dim_representation}
Like most dimension-reduction methods, PNC represents high-dimensional data through a set of low-dimensional coordinates, referred to as \textit{scores}. It also provides an alternative representation \textit{polar scores}, which offers a complementary geometric interpretation.

\textbf{Scores.}
\textit{PNC scores} are defined as scaled residuals. The score matrix is \( \boldsymbol{E} = [\boldsymbol{s}_1, \cdots, \boldsymbol{s}_d] \in \mathbb{R}^{n \times d} \). Residuals are computed sequentially from Step $1$ to Step $d$, but scores are arranged in reverse order in \( \boldsymbol{E} \) so that earlier columns correspond to more informative components. Under this convention, \(\boldsymbol{s}_1\) originates from Step $d$, and \(\boldsymbol{s}_d\) from Step $1$. More generally, \(\boldsymbol{s}_k\) corresponds to Stage \( d+1-k \).
PNC scores from Step \(k\) can be computed as
\begin{equation*}
	\label{eq:score}
	\boldsymbol{s}_{d+1-k} = \left( \prod_{i=1}^{k-1} \sin\alpha_i \right) \boldsymbol{\xi}_{d+1-k}, \qquad k = 1, \cdots, d .
\end{equation*}
Scaling factors \(\prod_{i=1}^{k-1} \sin \alpha_i\) depend on all preceding steps. They vary across steps but are constant across observations within each stage. Since \(\alpha_k \in (0, \frac{\pi}{2})\) for \(k = 1, \cdots , d-1\), \, all $\prod_{i=1}^{k-1} \sin(\alpha_i)$ are positive, and therefore scores inherit the same signs of the corresponding residuals. Although the scaling does not affect parameter estimation, it makes scores comparable across stages and facilitates the interpretation of their relative importance.

\textbf{Polar scores.}
Polar scores (\( \boldsymbol{s}_x, \boldsymbol{s}_y \)) are specifically introduced for the final step of PNC to address discontinuities in the standard scores near the boundary between positive and negative residual regions. As illustrated in Figure~\ref{fig:PNC}, two nearby observations, A and B, lie closely but on the opposite sides of this boundary between the positive and negative residual zones. Despite being close in the original space, they may receive standard PNC scores with opposite signs and large magnitudes, which can distort intuitive interpretations. To provide a continuous representation and reflect local proximity, we introduce polar scores as
\begin{equation*}
    \label{eq:polarsc}
    s_x = \mysize\, \cos(\beta) \quad \text{and} \quad
        s_y =\mysize\, \sin(\beta) , 
\end{equation*}  
where \(\beta \in [0, 2\pi)\) is the counterclockwise angle between \(\boldsymbol{v}_d\) and \(\boldsymbol{x}\). This transformation rotates $\boldsymbol{v}_{d}$ to the positive $x$-axis. 

Polar scores encode both size and relative position. In polar score plots (e.g., Figure~\ref{fig:simu1}), the radial distances directly represent the size of each observation, and the angular coordinate represents shape variation. 

Neither representation is universally superior. Standard PNC scores are better suited for capturing sequential variation across stages, whereas polar scores provide a more intuitive geometric representation at the final stage. In practice, considering both is often beneficial, as they offer complementary insights from different perspectives.
 
\subsection{Parameter estimation}
\label{sec:para_est}
As finding the sequence of nested cones underlies the dimension-reduction process, \textit{the key task} in PNC modeling is to estimate the sequence of cone parameters $ \boldsymbol{\Phi} =(\boldsymbol{v}_1, \dots, \boldsymbol{v}_d, \alpha_1, \dots, \alpha_{d-1}).$ 
The total number of parameters is
\begin{equation}
	\label{eq:dimPhi}
	\operatorname{dim}(\boldsymbol{\Phi}) = \sum_{k=1}^{d} (d + 2 - k) + (d - 1) = \frac{d(d + 5)}{2} - 1.
\end{equation}
For example, when \( d = 2 \), \(\boldsymbol{\Phi} = (\boldsymbol{v}_1, \boldsymbol{v}_2, \alpha_1)\), and \(\operatorname{dim}(\boldsymbol{\Phi}) = 6\).

Dimension reduction proceeds sequentially with parameters \( \boldsymbol{\Phi} \)  estimated at each stage using least squares. At Stage \(k\), the objective function (a specific form of Equation~\eqref{eq:genericobj}) is
\begin{equation*}
    \label{eq:objective}
    Q_{n,k}(\boldsymbol{v}_k, \alpha_k) = \sum_{i=1}^{n}{\xi}_{i,d + 1 - k}^2 \, , \end{equation*}
where ${\xi}_{i,d + 1 - k}$ is the residual of observation $i$ at Step \(k\). 
The least squares estimates are then given by
\(
    (\hat{\boldsymbol{v}}_k, \hat{\alpha}_k) = \arg \min_{(\boldsymbol{v}_k, \alpha_k)} Q_{n,k}(\boldsymbol{v}_k, \alpha_k) .
\) The parameters are constrainted by \(\boldsymbol{v}^T \boldsymbol{v} = 1\) and \( \alpha \in \left[ 0, \pi/2 \right] \) to ensure that \( \boldsymbol{v} \) is a unit vector and \(\alpha\) does not exceed that of a great cone. 
In practice, the optimization is performed numerically, e.g., using the \texttt{optim} function in \textsf{R} \citep{R}. The initial $\boldsymbol{v}$ is chosen as the mean direction (normalized to unit length), and the initial $\alpha$ as the average angular deviation from the initial $\boldsymbol{v}$.

The full estimation procedure of PNC parameters is summarized in Algorithm~\ref{alg:PNC}.

\begin{algorithm} [ht!]
	\caption{PNC parameter estimation}
	\label{alg:PNC}
	\begin{algorithmic}[0] 
		\STATE \textbf{Input:} Data matrix $\boldsymbol{X} \in \mathbb{R}^{(d+1)\times n}$

        \STATE \textbf{Initialize:} $\boldsymbol{X}^{(d)} \leftarrow \boldsymbol{X}$
        
		\FOR{\( k = 1, \ldots, d \)}
        
    		\STATE \textbf{(a) Parameter estimation:}
    		\STATE \hspace*{\algorithmicindent} Obtain $(\hat{\boldsymbol{v}}_{k}, \hat{\alpha}_k)$ by minimizing objective function $Q_{n,k}$

            \STATE \textbf{(b) Projection:}
            \STATE \hspace*{\algorithmicindent} Project the observations onto the fitted cone and compute their residuals and scores
            
            \STATE \textbf{(c) Dimension reduction:}
    		\STATE \hspace*{\algorithmicindent} Map \( \boldsymbol{X} ^ {(d + 1 - k)} \) to lower dimension: 	$\boldsymbol{X} ^ {(d - k)} = \operatorname{M} \left( \boldsymbol{X} ^ {(d + 1 - k)}, \hat{\boldsymbol{v}}_{k} \right)$
                
        
		\ENDFOR
        
        
		\STATE \textbf{Output:} PNC score matrix \(\boldsymbol{E}\) and parameter collection \(\boldsymbol{\Phi}\)
	\end{algorithmic}
\end{algorithm}

\subsection{Back-fitting}
\label{sec:bf}
PNC can be inverted to reconstruct the original data, which we denote as \( \operatorname{PNC}^{-1} \). Given scores, sizes, and parameters, \( \operatorname{PNC}^{-1} \) can recover the original data matrix $\boldsymbol{X}$ through a sequential procedure from Step $d$ to Step $1$.
For an observation \(\boldsymbol{x}\) with residual vector \( \boldsymbol{\xi} \), let \( \boldsymbol{y}^{(d + 1 - k)} \) denote the reconstructed vector after inverting Step \( k \). The inversion begins at Step \(d\), which can be recovered by 
\begin{equation*}
    \boldsymbol{y}^{(1)} = \begin{bmatrix}
					\cos \left(\phi + \frac{\xi_1}{\mysize} \right)\, \mysize \\ 
                    \sin \left(\phi + \frac{\xi_1}{\mysize} \right)\, \mysize
                    \end{bmatrix}, 
    \label{eq:bf1}
\end{equation*}
where \(\phi \in [0, 2\pi)\) is the positive angle between \(\boldsymbol{v}_d\) and \((1, 0)\), and \( (\phi + \xi_1 / \mysize) \) is therefore the angle between \(\boldsymbol{x}\) and \((1, 0)\).
For $k = d-1, \dots, 1$, the \(k^{\text{th}}\) step can be inverted by
	      \begin{equation*}
		      \boldsymbol{y}^{(d+1-k)}
		      = \boldsymbol{R}_{k}^T \begin{bmatrix*}[l]
			      \sin \left(\alpha_{k} + \frac{\xi_{d+ 1 - k}}{\mysize} \right)\, \boldsymbol{y}^{(d - k)} \\ 
                  \cos \left(\alpha_{k} + \frac{\xi_{d+ 1 - k}}{\mysize}\right)\, \mysize
		      \end{bmatrix*},
            \label{eq:bfk}
	      \end{equation*}
where \(\boldsymbol{R}_{k}\) is a rotation matrix of \(\boldsymbol{v}_{k}\). 

The above formulas are derived using $\xi_r$. For chordal residuals $\xi_c$, replace $\xi/\mysize$ with $2\arcsin\!\left[\xi/(2\mysize)\right]$. The full PNC back-fitting procedure is summarized in Algorithm~\ref{alg:PNCbackfit}.

Note that when the full score vector \(\boldsymbol{s}\) is used, the observation can be reconstructed exactly, up to numerical error, i.e., \(\boldsymbol{x}^{(d)} = \boldsymbol{y}^{(d)}\). When scores from earlier stages are omitted, the back-fitting procedure can still yield a good approximation because early-stage scores typically carry less information. The choice of the number of scores to retain depends on the data characteristics. Contribution of scores from Step~1 through Step~\(k\) can be removed by setting corresponding scores to zero.

A special case arises when all scores are set to zero, and only the sizes are retained. The resulting back-fitting represents the mean shape at the sizes. An even more specific case occurs when all scores are set to zero, and the size is fixed at the sample mean size. The resulting reconstruction corresponds to the sample PNC mean size-and-shape. This means size-and-shape differs from the arithmetic mean size-and-shape both in concept and in value.

\begin{algorithm} [ht!]
	\caption{PNC back-fitting}
	\label{alg:PNCbackfit}
	\begin{algorithmic}[0]
		\STATE \textbf{Input:} Scores $\boldsymbol{E}$, sizes $\boldsymbol{r}$, and parameter collection $ \boldsymbol{\Phi} $
        
		\STATE \textbf{(a) Recover residuals:} Compute residuals $\boldsymbol{\Xi}$ from scores $\boldsymbol{E}$ 
        \STATE \textbf{(b) Backfit the last stage and compute $\boldsymbol{Y}^{(1)}$}

        \STATE \textbf{(c) Sequential backfitting:}
            \STATE \hspace*{\algorithmicindent} \textbf{for} $k = d - 1, \dots, 1$ \textbf{do}
            \STATE \hspace*{2\algorithmicindent} Compute $\boldsymbol{Y}^{(d + 1 - k)}$ based on $\boldsymbol{Y}^{(d - k)}$
            \STATE \hspace*{\algorithmicindent} \textbf{end for}    
		\STATE \textbf{Output: } Reconstructed data matrix \( \boldsymbol{Y} \)
	\end{algorithmic}
\end{algorithm}

\subsection{Fast approximation of PNC}
\label{sec:fastPNC}

Fast approximation of PNC (fast PNC) extends the standard PNC framework to ultra-high-dimensional settings, where direct estimation becomes computationally expensive. As shown in Equation~\eqref{eq:dimPhi}, the number of parameters grows quadratically with \(d\), which can become prohibitive when $d$ is large. 
To address this limitation, we introduce fast PNC, which first reduces the dimensionality of the input data using a PCA-based transformation \( \operatorname{T}_\text{PCA}\), and then applies standard PNC to the reduced representation. This approach follows a similar idea of fast PNS \citep{Drydenetal19,Monemetal25}. Formally,
\(
    \operatorname{PNC}_\text{fast}(\boldsymbol{X}, p) = \operatorname{PNC} \left(\operatorname{T}_\text{PCA}(\boldsymbol{X}, p) \right),
\)
where \(\operatorname{T}_\text{PCA}(\boldsymbol{X}, p)\) produces a $(p + 1) \times n$ data matrix.

The number of retained principal components $p$ satisfies \( p \le p_{\max} = \min \left\{ d + 1, \, n - 1 \right\}\). 
When $d \gg n$, $p_{\text{max}}$ is considerably smaller than $d$, leading to substantial dimensional reduction and computational savings. When $p = p_{\max}$,  no information is lost in the PCA-based transformation step.

PCA-based transformation can be summarized as a sequence of space transformations with three stages (see Figure S1 in the supplementary material for a graphical illustration): 
\[
\mathcal{S}^d \times \mathbb{R}^+
\;\longrightarrow\;
\mathcal{T}_{\bar{\boldsymbol{x}}}(\mathcal{S}^d) \times \mathbb{R}^+
\;\longrightarrow\;
\mathcal{T}_{\bar{\boldsymbol{x}}}(\mathcal{S}^p) \times \mathbb{R}^+
\;\longrightarrow\;
\mathcal{S}^p \times \mathbb{R}^+.
\]

\textbf{Step 1: Mapping to the tangent space.}
As PCA performs in the tangent space, all observations are mapped to the tangent space at a reference point \(\bar{\boldsymbol{x}} \in \mathcal{S}^{d}\), which is a unit vector and represents the normalized mean direction:
\begin{equation*}		      
			\bar{\boldsymbol{x}}  = \frac{\frac{1}{n} \sum_{j = 1} ^ n \frac{\boldsymbol{x}_j}{\mysize_j}}{\left\lVert \frac{1}{n} \sum_{j = 1} ^ n \frac{\boldsymbol{x}_j}{\mysize_j} \right\rVert} .
\end{equation*}
The tangent vector for observation \(\boldsymbol{x}\) in the space \( \mathcal{T}_{\bar{\boldsymbol{x}}} (\mathcal{S}^{d}) \times \mathbb{R} ^+\) is then
\(
    \boldsymbol{t} = \boldsymbol{x} - (\bar{\boldsymbol{x}}^T \boldsymbol{x}) \bar{\boldsymbol{x}}
\). 
Ordinary PCA is then applied to the matrix \(\boldsymbol{T} = [\boldsymbol{t}_1, \cdots \boldsymbol{t}_n]\). Let $\boldsymbol{V}_j$ denote principal directions and 
$\tilde{\boldsymbol{q}}$ the corresponding PCA scores. To preserve size information, the PCA scores are rescaled as
$
   \boldsymbol{q} = \frac{\manDist( \boldsymbol{x}, \bar{\boldsymbol{x}}) \, \mysize}{\left\lVert \boldsymbol{t} \right\rVert} \tilde{\boldsymbol{q}}.$

\textbf{Step 2: Dimensionality reduction.}
The reduced representation in \( \mathcal{T}_{\bar{\boldsymbol{x}}} (\mathcal{S}^{p} \times \mathbb{R} ^+)\) is
$
    \boldsymbol{u} = \sum_{j=1}^{p} q_{j} \boldsymbol{V}_j .
$

\textbf{Step 3: Mapping back to manifold.} 
The reduced representations are mapped back from the tangent space to a manifold via the exponential map
\begin{align*}
    \tilde{\boldsymbol{x}}  = \cos \left(\frac{\left\lVert \boldsymbol{u} \right\rVert}{\mysize} \right)\, \mysize\, \bar{\boldsymbol{x}} +
    \sin \left(\frac{\left\lVert \boldsymbol{u} \right\rVert}{\mysize}\right)\, \frac{\mysize}{\left\lVert \boldsymbol{u} \right\rVert}\, \sum_{j=1}^{p} q_{j}\boldsymbol{V}_j .
\end{align*}
Here \(\bar{\boldsymbol{x}}\) and \(\boldsymbol{V}\) are orthogonal and together span the \( (p+1) \)-dimensional space, the coordinates of \(\tilde{\boldsymbol{x}}\) are therefore
\begin{equation*}
    \tilde{\boldsymbol{x}} = \left[\cos\left(\frac{\left\lVert \boldsymbol{u} \right\rVert}{\mysize} \right)\, \mysize, \;
    \sin \left(\frac{\left\lVert \boldsymbol{u} \right\rVert}{\mysize} \right)\, \frac{\mysize}{\left\lVert \boldsymbol{u} \right\rVert}\, q_1, \;
    \cdots, \;
    \sin \left(\frac{\left\lVert \boldsymbol{u} \right\rVert}{\mysize} \right)\, \frac{\mysize}{\left\lVert \boldsymbol{u} \right\rVert}\, q_p \right]^T  \in \mathcal{S}^{p} \times \mathbb{R} ^+ / \sim.
\end{equation*}

The steps reduce each observation from dimension $(d+1)$ to $(p+1)$, after which standard PNC is applied.
Since fast PNC consists of a PCA-based transformation followed by standard PNC, its backfitting is obtained first by inverting standard PNC and then \(\operatorname{T}_\text{PCA}\):
\begin{equation*}
    \operatorname{PNC}_\text{fast}^{-1} ( \boldsymbol{E}, \boldsymbol{\Phi}, \boldsymbol{V} ) =\operatorname{T}_\text{PCA}^{-1} \left\{ \operatorname{PNC}^{-1}(\boldsymbol{E}, \boldsymbol{\Phi}), \boldsymbol{V} \right\},
\end{equation*}
where \( \operatorname{PNC}^{-1}(\boldsymbol{E}, \boldsymbol{\Phi}) \) restores a matrix \(\boldsymbol{G}\) with row vectors \(\boldsymbol{g}_1, \cdots, \boldsymbol{g}_{p + 1}\).
The inverse PCA transformation does not need to invert every step of \(\operatorname{T}_\text{PCA}\). Instead, it can be computed as
\begin{equation*}
	\label{eq:PCAbackfit}
	\boldsymbol{Y}^{(d)} = \bar{\boldsymbol{x}} \boldsymbol{g}_1 + \sum_{j=1}^{p} \boldsymbol{V}_j \boldsymbol{g}_{j+1} .
\end{equation*}

\subsection{Properties}
\label{sec:properties}
The PNC fitting procedure described in Section~\ref{sec:para_est} minimizes the sum of squared residuals at each stage. Therefore, it belongs to the class of $M$-estimators. Under standard regularity assumptions, we have the following consistency and asymptotic normality results.

\textbf{Consistency.}
Assume that: (i) observations are independent and identically distributed; (ii) $\boldsymbol{\Phi}$ is identifiable, in the sense that the population objective function $Q(\boldsymbol{\Phi})$ has a unique minimizer at the true $\boldsymbol{\Phi}$; and (iii) residuals have finite second moments and are well-defined at each stage.
Under these conditions, a uniform law of large numbers applies to $Q_n(\boldsymbol{\Phi})$, where $Q_n(\boldsymbol{\Phi})$ is the empirical objective function for a sample of size $n$. For an estimator obtained from a sample of size $n$, $\hat{\boldsymbol{\Phi}}_n$, we have the consistency of standard $M$-estimation such that
\(
\hat{\boldsymbol{\Phi}}_n \rightarrow \boldsymbol{\Phi}\)\, as $n \rightarrow \infty.
$
\textbf{Asymptotic normality.}
If, in addition, the residual functions $\boldsymbol{\xi}_{k}(\boldsymbol{x};\boldsymbol{\Phi})$ are sufficiently smooth in a neighborhood of the true value, and the Hessian matrix of $Q(\boldsymbol{\Phi})$ is nonsingular at the true value, then the standard M-estimation theory gives \,\(
\sqrt{n} \left(\hat{\boldsymbol{\Phi}}_n - \boldsymbol{\Phi} \right) \xrightarrow{d} \mathcal{N}(0, \boldsymbol{\Sigma}).
\)

Estimation variability propagates across stages, because PNC estimation proceeds sequentially across levels of nested cones. In practice, uncertainty can be assessed using a bootstrap that resamples observations and refits the full PNC model, thereby capturing stagewise dependence. We illustrate this process in Section~\ref{sec:simuconvergence}.

\textbf{Remarks.}
Such asymptotic properties established for spherical PNC extend to more general manifold settings (as described in Section~\ref{sec:conespace}), provided that: (i) $M$ is smooth and compact; (ii) the residual functions are differentiable with respect to parameters; (iii) the geodesic projections $\pi_k$ are locally well-defined, and (iv) the parameters are identifiable. Under such conditions, stagewise fittings remain an $M$-estimation problem, and consistency and asymptotic normality follow from standard arguments.

\section{Simulation study}
\subsection{Comparison between PNC, PCA, and PNS}

To illustrate how PCA, PNS, and PNC capture the underlying structure of data, we conduct a numerical experiment in the space \(\mathcal{S}^2 \times \mathbb{R}^+ / \sim\). The low-dimensional setting is chosen so that both the original data and the dimension-reduced representations can be visualized, thereby providing intuition for the behavior of each method.

\begin{figure}[t!]
    \centering
    \includegraphics[width=.9\linewidth]{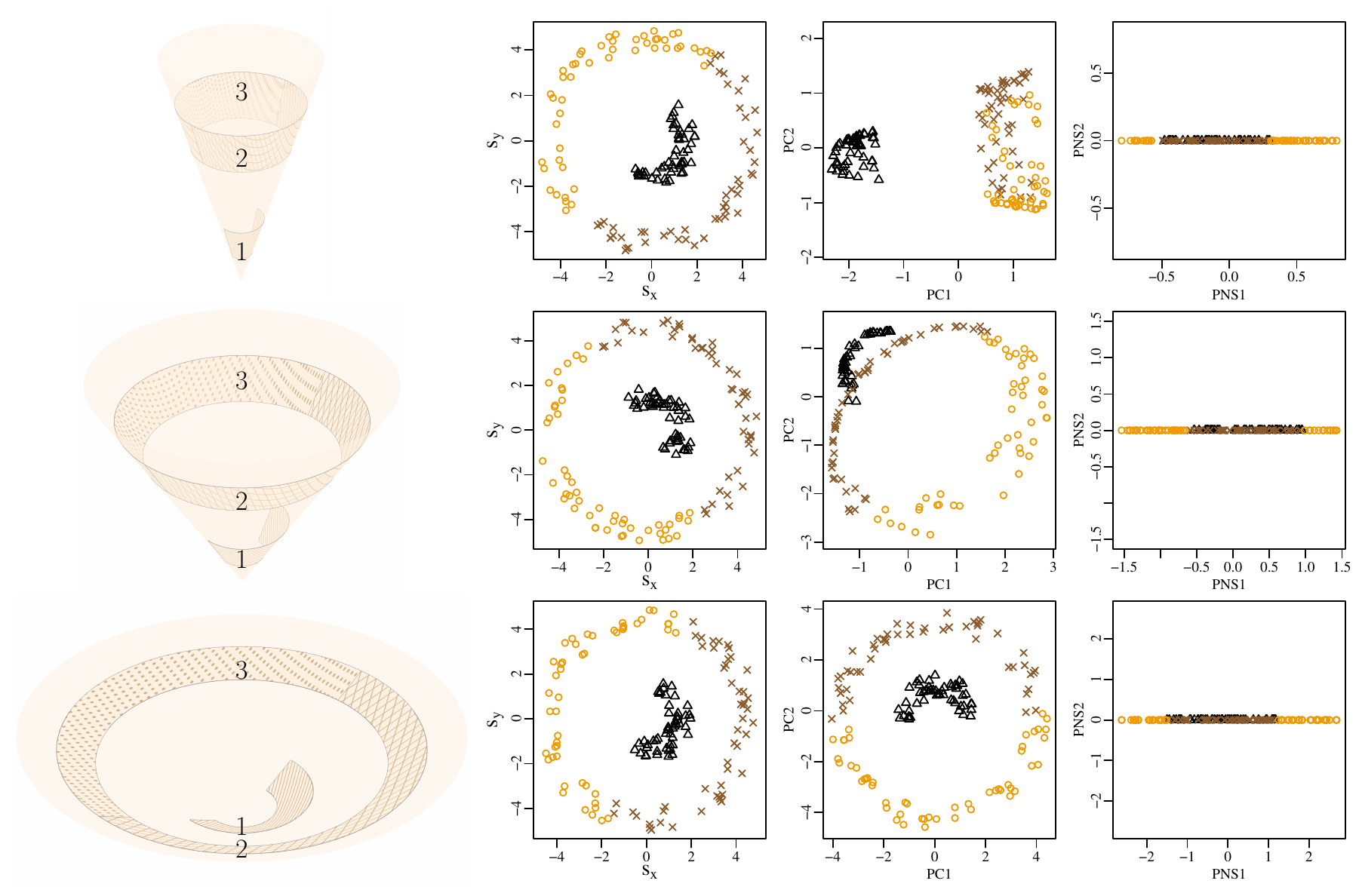}
    \caption{
        Simulation setup and results.
        Rows correspond to $\alpha = \pi/12$, $\alpha = \pi/6$, and $\alpha = \pi/3$ (top to bottom). 
        Columns show (from left to right):  original data sampling, PNC polar scores, PCA scores, and PNS scores.
        Sample regions are defined by radial and angular ranges:
        Region~1 ($\triangle$): $\mysize \in [1,2]$, $\theta \in [0,\pi]$ ;
        Region~2 ($\times$): $\mysize \in [4,5]$, $\theta \in [0,\pi]$;
        Region~3 ($\circ$): $\mysize \in [4,5]$, $\theta \in [\pi,2\pi]$.
        }
    \label{fig:simu1}
\end{figure}

The data are sampled uniformly and independently from three distinct regions on a conical surface, differing in angular position and radial distance (which are sampled independently). Specifically, we consider two radial intervals $[1,2]$ and $[4,5]$, two angular intervals $[0,\pi]$ and $[\pi, 2\pi]$, and three opening angles $\alpha$: $\pi / 12$, $\pi / 6$, and $\pi / 3$, as shown in the left column of Figure~\ref{fig:simu1}. The parameter $\alpha$ controls the curvature of the underlying manifold: smaller values correspond to a narrow cone, while larger $\alpha$ produces a flatter structure.
The results are shown in Figure~\ref{fig:simu1}.
PNC effectively captures both clustering and proximity relationships in the data across all simulation settings. In particular, as shown in the second column of Figure~\ref{fig:simu1}, the polar score representation preserves both radial (size) and angular (shape) variation, leading to a clear separation of the three regions.
In contrast, the performance of PCA depends strongly on \(\alpha\), as shown in the third column of Figure~\ref{fig:simu1}. When \(\alpha\) is small, PCA primarily captures radial (size) variation, and the first principal component tends to align with the cone axis. As \(\alpha\) increases, angular variation becomes more prominent, and the principal directions shift accordingly. This observation reflects the fact that PCA identifies directions of maximal variance while overlooking the intrinsic geometry of the data.
PNS, which operates on data normalized to unit size, removes size information entirely. As shown in the last column of Figure~\ref{fig:simu1}, observations from different radial regions become indistinguishable, which again leads to a loss of structure in the representation.

A spiral example on the cone surface is provided in the supplementary material (Figure~S2), which shows similar behavior.


\subsection{Back-fitting efficiency}
\label{sec:back-fitting simu}
We also evaluate the representation efficiency and reconstruction performance of PNC, PNS, and PCA using back-fitting distance and variance explained. The results are reported in the supplementary material (Section~S4, Figure~S3) and show that PNC consistently achieves smaller reconstruction error and explains a larger proportion of variance than PCA, with the largest improvements at moderate cone curvature.

\subsection{Convergence of parameter estimation with sample sizes}
\label{sec:simuconvergence}
To assess convergence to true parameters and estimation uncertainty, we study the behavior of PNC parameter estimates as the sample size increases. We consider hypercones in \(\mathcal{S}^3 \times \mathbb{R}^+ / \sim\), which provide a parameter space that is rich enough to reveal general behavior while low-dimensional enough to visualize individual estimates.
Data are generated via the back-fitting procedure described in Section~\ref{sec:bf}, using parameters $\boldsymbol{\Phi}$ in Table~\ref{tab:sim-params}. 

Bootstrap resampling is used to assess estimation uncertainty. For each sample size $n$, an independent dataset is generated, from which 1{,}000 bootstrap samples are drawn with replacement. PNC parameters are re-estimated for each bootstrap sample, and empirical 90\% confidence intervals are constructed. This entire procedure is repeated 100 times for each $n$, and the resulting confidence‑interval widths are averaged across repetitions.

Directions are unit vectors whose components satisfy the constraint \(\boldsymbol{v}^T \boldsymbol{v} = 1\). This dependence among components makes it inappropriate to construct confidence intervals for each coordinate independently. To obtain an unconstrained parameterization, we instead express $\boldsymbol{v}$ using hyperspherical coordinates. For a unit vector $\boldsymbol{v}$ with $m$ components, the mapping $h : \boldsymbol{\theta} \mapsto \boldsymbol{v}$ is defined as
\begin{equation*}
v_k =
\begin{cases}
\Big(\prod_{j=1}^{k-1}\sin\theta_j\Big)\cos\theta_k,
& k=1,\dots,m-1,\\[6pt]
\prod_{j=1}^{m-1}\sin\theta_j,
& k=m.
\end{cases}
\label{eq:hypersphere}
\end{equation*}
where \(\theta_j \in [0, \pi]\) for \(j = 1, \ldots, m-2\) and \(\theta_{m-1} \in [0, 2\pi]\).
Under this parameterization,  the original parameters \( \{ \boldsymbol{v}_1, \boldsymbol{v}_2, \boldsymbol{v}_3, \alpha_1, \alpha_2 \} \) are transformed into \( \{ \theta_{11}, \theta_{12}, \theta_{13}, \theta_{21}, \theta_{22}, \theta_{31}, \alpha_1, \alpha_2 \} \), where \(\boldsymbol{v}_1 = h(\theta_{11}, \theta_{12}, \theta_{13})\), \(\boldsymbol{v}_2 = h(\theta_{21}, \theta_{22})\), and \(\boldsymbol{v}_3 = h( \theta_{31})\).

\begin{table}[ht]
\centering
\caption{Parameters used for data generation}
\label{tab:sim-params}

\begingroup
\renewcommand{\arraystretch}{0.8}
\begin{tabular}{ll ll}
\toprule
Param. & Value & Param. & Value \\
\midrule
$\boldsymbol{v}_1$ & $(0.5,0.5,0.5,0.5)^\top$ &
$\mysize$ & $ \;\mathcal{U}(10,20)\;$ \\

$\boldsymbol{v}_2$ & $\left(1/\sqrt{3},\,1/\sqrt{3},\,1/\sqrt{3}\right)^\top$ &
$\boldsymbol{\xi}_1$ & $\displaystyle \mathcal{N}\!\left(0,\left(\pi \mysize / 3 \right)^2\right)$, truncated to $[-\pi\mysize,\;\pi\mysize]$\\

$\boldsymbol{v}_3$ & $\left(1/\sqrt{2},\,1/\sqrt{2}\right)^\top$ &
$\boldsymbol{\xi}_2$ & $\displaystyle \mathcal{N}\!\left(0,1\right)$, truncated to $\left[-\alpha_2 \, \mysize,\;\alpha_2 \, \mysize\right]$\\
$(\alpha_1,\alpha_2)$ & $\left(\pi / 6,\, \pi / 4 \right)$ &
$\boldsymbol{\xi}_3$ & $\mathcal{N}(0,0.3^2)$, truncated to $\left[-\alpha_1 \, \mysize,\;\alpha_1 \, \mysize\right]$ \\
\bottomrule
\end{tabular}
\endgroup
\end{table}

\begin{figure}[ht!]
	\centering	
    \includegraphics[width=1\linewidth]{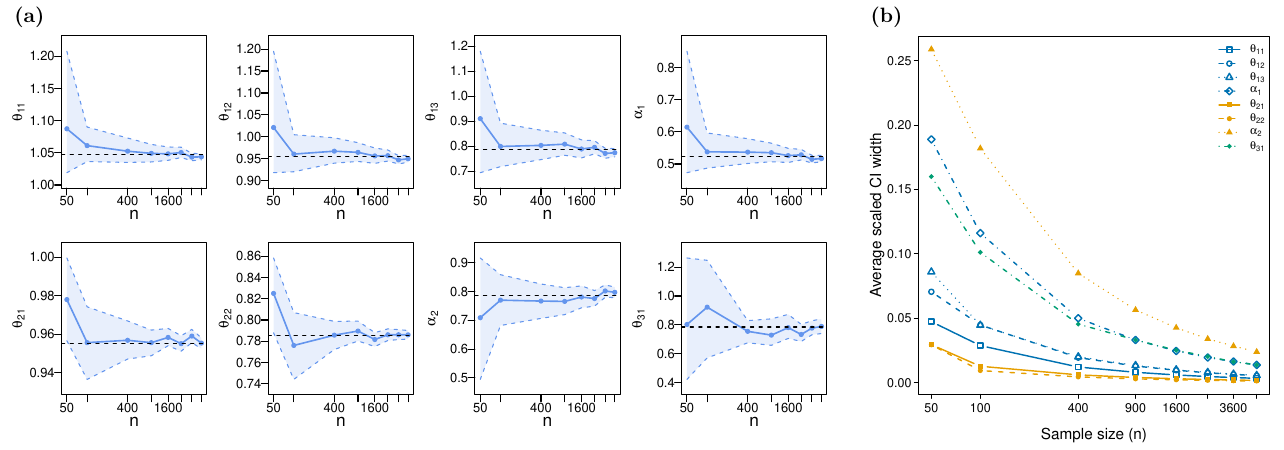}
	\caption{(a) PNC parameter estimates as a function of sample size \(n\), and $n \in \{50, 100, 400, 900, 1600, 2500, 3600, 5000\}$. Solid blue lines represent mean estimates, shaded bands represent 90\% confidence intervals, and dashed lines indicate true parameter values. 
    (b) Average normalized confidence interval widths versus sample size $n$ (on a \(\log_{10}\) scale).
    }
    \label{fig:ci_bs}
\end{figure}

Figure~\ref{fig:ci_bs}~(a) shows representative results for parameter estimation, including 90\% confidence intervals, along with the mean estimates and true value references. As the sample size $n$ increases, the mean estimates of all parameters approach the true parameter values, indicating consistency of the proposed estimator. At the same time, the confidence intervals become narrower as \(n\) grows, reflecting reduced estimation variability.


Figure~\ref{fig:ci_bs}~(b) shows the average confidence interval widths across all parameters. As the parameters have different supports (e.g.,  \( [0,\pi/2] \), \( [0,\pi] \), and \( [0,2\pi] \)), the interval widths are normalized by their respective ranges to allow comparison on a common scale. The \(x\)-axis is the sample size $n$ shown on a \(\log_{10}\) scale. The results show a clear decrease in confidence interval width as \(n\) increases. This behavior provides empirical evidence for the consistency and asymptotic properties of PNC, described in Section~\ref{sec:properties}.

\section{Real data analysis}
This section demonstrates the application of PNC to real datasets. The crab and rat skull datasets illustrate the standard PNC framework, while the molecule dataset demonstrates the fast PNC approach for ultra-high-dimensional shape data.

\subsection{Crabs data}
The crabs dataset \citep{campbell1974multivariate}  consists of five morphological measurements recorded on 200 rock crabs of the genus {\it Leptograpsus}, corresponding to 50 crabs from each combination of two species and two sexes. The dataset is available in the {\tt MASS} package of R \citep{R}.
The variables are: \texttt{FL} (frontal lobe size in mm), \texttt{RW} (rear width in mm), \texttt{CL} (carapace length in mm), \texttt{CW} (carapace width in mm), and \texttt{BD} (body depth in mm). These measurements form the data matrix $\boldsymbol{X}  \in \mathbb{R}^{5 \times 200} $. The dataset also includes two categorical variables: \texttt{sp} (species: \texttt{B} for blue or \texttt{O} for orange) and \texttt{sex}. These two variables define four groups, namely: blue male (B.M), blue female (B.F), orange male (O.M), and orange female (O.F).
This dataset is a classical benchmark in multivariate analysis and is well-suited for illustrating joint size-and-shape analysis, as the measurements capture both overall size differences and shape variation across groups. Since the measurement scales are meaningful, no centering or scaling is applied.

\begin{figure}[ht!]
    \centering
    \includegraphics[width=1\linewidth]{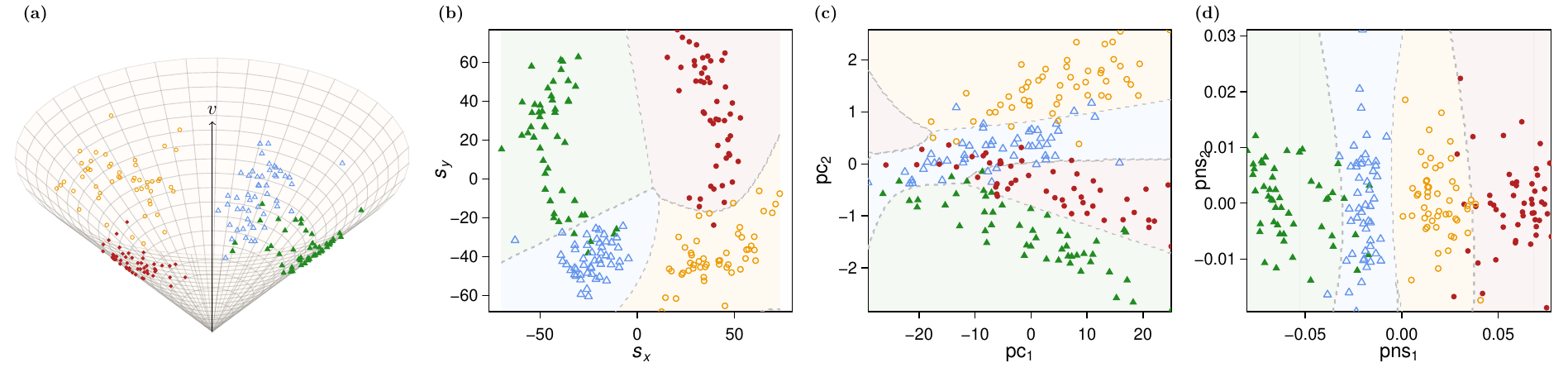}
    \caption{(a) Crabs projected on the fitted cone. 
    (b–d) 2D representations: (b) PNC polar scores, (c) PCA scores, and (d) PNS scores, with QDA decision boundaries. 
    Symbols and colors indicate groups: B.M. (green $\blacktriangle$), B.F. (blue $\triangle$), O.M. (red $\bullet$), and O.F. (orange $\circ$).}
    \label{fig:crabComp}
\end{figure}
The crabs projected on the fitted 3D cone are shown in Figure~\ref{fig:crabComp}~(a). Each point represents a crab observation, and its position reflects both size (radial direction) and shape (angular position).
The cone representation provides a natural visualization of proximity relationships. Observations from the same species or sex tend to cluster together, while distances between clusters reflect differences in both size and shape. In particular, male and female crabs are clearly separated along one direction, while species differences are presented along another direction. This geometric representation provided by PNC preserves neighborhood structure on the manifold, which is not guaranteed by linear models and is not accurately captured by PNS due to its removal of size information. The corresponding measure of similarity is given by the hypercone geodesic distance, as detailed in Section~\ref{sec:conegeo}. 


As shown in Figure~\ref{fig:crabComp}~(b), the polar score representation of PNC (introduced in Section~\ref{sec:low-dim_representation}) reveals clear structure in the data and yields clearer group separation, reflected by the quadratic discriminant analysis (QDA; \citealp{venables2013modern}) decision boundaries. Separation by sex occurs primarily along the vertical axis, with male crabs clustering in the upper region and female crabs in the lower region. Species differences are captured mainly along the horizontal direction, distinguishing blue and orange crabs. Importantly, radial distance in the polar plot corresponds directly to size, while angular position reflects shape variation. This provides a direct and interpretable visualization of joint size-and-shape structure. Polar score plots also convey much of the information contained in a 3D visualization by unfolding a three-dimensional surface into a two-dimensional display.

In contrast, PCA (panel c) produces a sector-like pattern in which groups overlap substantially, making separation less clear. This reflects the limitation of linear projections in capturing curved manifold structure.
PNS (panel d) shows some clustering, but O.M and O.F are less well separated. This is because the normalization in PNS removes size information,  reducing its ability to distinguish groups that differ primarily in scale. 




\subsection{Rat skulls data}
The rat skulls dataset \citep{bookstein1997morphometric} consists of two-dimensional landmarks collected from 21 individuals at eight time points (7, 14, 21, 30, 40, 60, 90, and 150 days of age). Each observation contains eight landmarks. After excluding four incomplete observations with missing landmarks, 164 samples remain. The configurations are first aligned using generalized Procrustes analysis \citep{dryden2016statistical,Rshapes}, and then analyzed using PNC and PNS.

\begin{figure}[ht!]
    \centering    
    \includegraphics[width=1\linewidth]{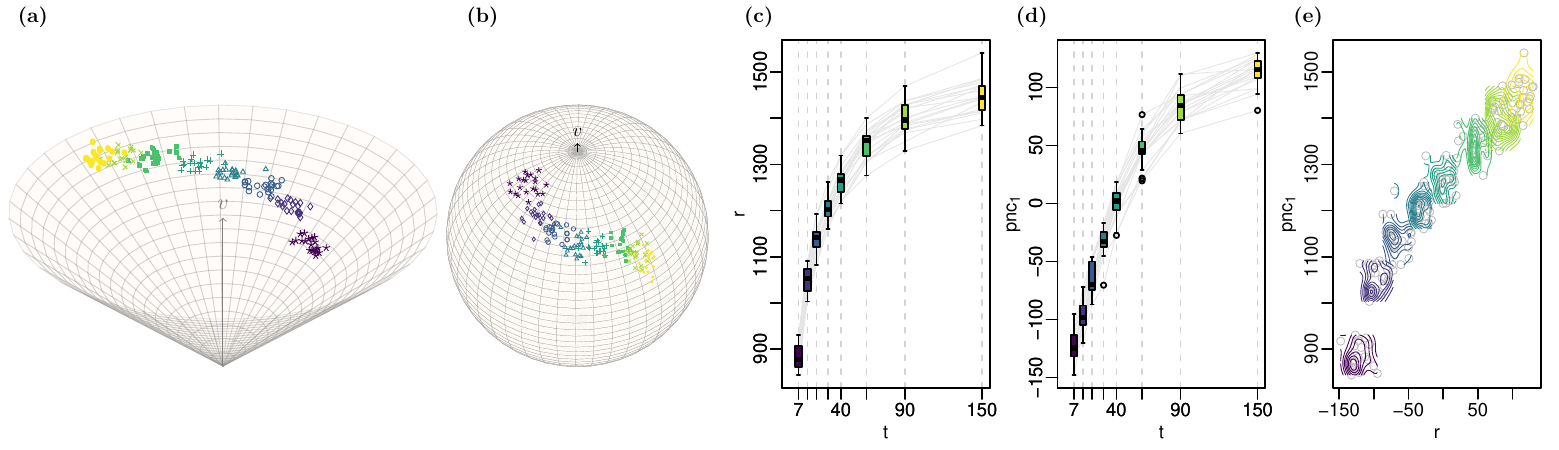}
    \caption{Rat skulls on the PNC cone (a) and the PNS unit sphere (b), with longitudinal changes in PNC size (c), PNC score~1 (d), and their joint distribution (e). Colors indicate time progression from early (dark purple) to late (yellow) across all panels.
    In (a) and (b), markers denote scan times: $\star$ (7 days), $\diamond$ (14 days), $\circ$ (21 days), $\triangle$ (30 days), $+$ (40 days), $\blacksquare$ (60 days), $\times$ (90 days), and $\bullet$ (150 days). In (c) and (d), gray curves show individual trajectories, with boxplots summarizing distributions at each time point. In (e), gray points represent observations, and contours show the estimated joint density.}
    \label{fig:rat}
\end{figure}

The results are presented in Figure~\ref{fig:rat}. In (a) and (b), observations exhibit a clear temporal progression along the manifold, with points moving smoothly from early (purple) to late time points (yellow). This pattern reflects a strong temporal structure in the data. In addition, observations cluster locally, indicating temporal consistency and low variability within local temporal ranges. 
However, the two methods differ in how this progression is represented. On the PNC cone in (a), the points are arranged along a trajectory that incorporates both radial and angular variation. The radial direction captures size increase, while angular variation reflects shape changes. 
In contrast, the PNS sphere in (b) preserves temporal ordering but compresses variation onto the unit sphere, emphasizing shape but not size, and thus only partially capturing growth dynamics.

Figure~\ref{fig:rat} (c - e) provides a more explicit view of temporal variation. Panel~(c) shows the evolution of size over time. A clear increasing trend is observed, with rapid growth in early stages followed by slower growth at later times, which results in a concave trajectory. Panel~(d) shows PNC score 1, which also increases with time and shows a similar trend. Compared with size, the increase in PNC score~1 is more gradual, which suggests that shape changes continue even after size growth slows.
Panel~(e) shows the joint distribution of size and PNC score~1. The contours also illustrate a continuous trajectory over time. Joint distributions at later time points (90 and 150 days) are less well separated compared to earlier stages. It reflects reduced variability as the shape and size stabilize during maturation.
Overall, PNC provides a coherent representation of the growth process and captures both size increase and shape evolution in a unified framework. 

\subsection{Molecule data}
\label{sec:moldata}
The dataset consists of seven molecular configurations derived from cryo-electron microscopy (cryo-EM) data in the Worldwide Protein Data Bank \citep{berman2003announcing}. Each configuration comprises 2{,}498 three-dimensional atoms and represents the configuration of a multi-protein complex called the sliding clamp that forms a ring around a section of DNA, and is part of the cellular machinery that makes DNA replication happen. 
Clamp loaders are proteins that bind to the clamp and switch it between `open' and `closed' states so it can bind and unbind from the DNA. 
Cryo-EM studies of the complex between the clamp and clamp loader have been able to detect a range of states between `open' and `closed', and our aim is to visualize a path between open and closed states to show the dynamics of the process.

\begin{figure}[ht!]
\centering
\includegraphics[width=1\linewidth]{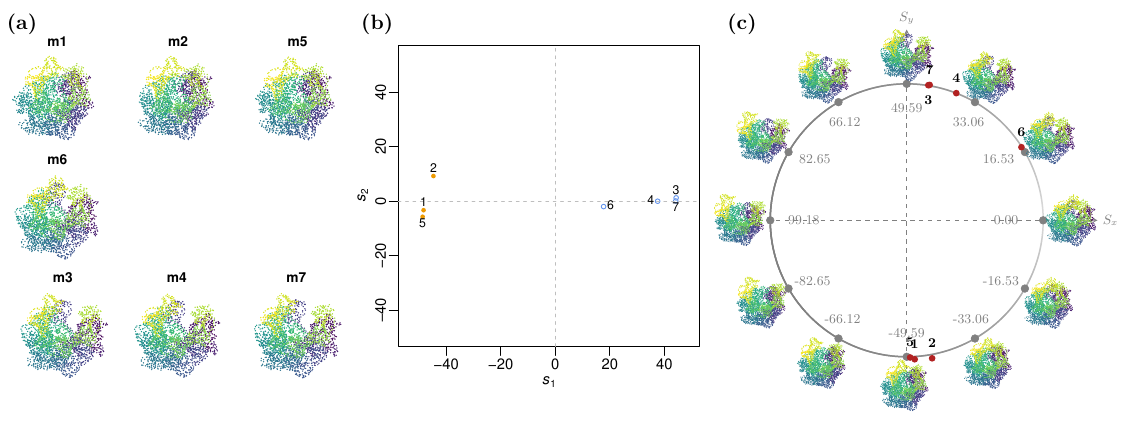}
\caption{(a) Seven molecules represented by 2,498 3D landmarks. Colors indicate landmark ordering and correspondence across molecules.
(b) PNC scores: solid circles ($\bullet$): visually closed molecules; open circles ($\circ$): open ones.
(c) Reconstructed configurations along the polar score orbit. Back-fitted configurations are displayed at given score values. Bold labels indicate the original observations.
}
\label{fig:mol}
\end{figure}

Since shapes of the multi-protein complexes are invariant to translation and rotation, each configuration is mean-centered and aligned by generalized Procrustes analysis \citep{dryden2016statistical,Rshapes}. 
The aligned configurations are shown in Figure~\ref{fig:mol}~(a). Visual inspection suggests two primary conformational states: samples m1, m2, and m5 exhibit closed postures, whereas m3, m4, and m7 show open postures. Sample m6 lies between these two groups and appears closer to the open configuration.

Each configuration is vectorized into a $(2498 \times 3)$-dimensional vector, resulting in an ambient dimension of $d = 7494 - 6 = 7488$ (with three dimensions removed due to each of translation and rotation). Fast PNC is applied with
\(p = p_{\max} = n - 1 = 6\).
The resulting PNC scores are shown in Figure~\ref{fig:mol}~(b).  
PNC score~1 provides a clear separation between closed and open configurations and explains more than 98\% of the total variance.
Panel~(c) shows a similar separation with the polar score representation. In both representations, sample m6 lies between the two groups and closer to the open group, consistent with its intermediate conformation visually observed in the original data.

Back-fitting is used to map PNC scores back to molecular configurations. Using the mean molecular size and a PNC score~1 range of \([-99.18, 99.18]\) (approximately $\pm 2$ standard deviations), we reconstruct a continuous trajectory of molecular shapes along the circular orbit in Figure~\ref{fig:mol}~(c), where representative configurations are displayed. As the score varies along the trajectory, the molecular posture undergoes a smooth transition from a closed to an open configuration. Starting at the bottom of the orbit, the structure gradually opens, reaches a maximally open state near the top, and then returns toward a closed posture.

The original samples align well with this trajectory: m1, m2, and m5 lie near the lower part of the orbit (closed configurations), whereas m3, m4, and m7 lie near the upper region (open configurations). Sample m6 lies between these groups, in the upper-right region of the orbit, indicating its intermediate state closer to the open group. This trajectory is constructed using seven molecules, and including more configurations would further refine it. A video of the smooth motion is provided in the supplementary material.

\section{Discussion}
We have proposed the PNC framework as a geometric approach for cone data dimension reduction, which is particularly useful for joint size-and-shape dimension reduction. Conceptually, it generalizes Principal Nested Spheres \citep{Jungetal12} to broader settings by retaining size throughout the analysis and replacing nested spheres with nested hypercones. 
The estimation of PNC up to a scale can be computed as a weighted version of PNS using the objective function (\ref{eq:genericobj}), where the weighting of the squared residual of each point is by its squared size. 
In the special case where observations are normalized to unit size, the PNC framework reduces to PNS. 

Compared with existing dimension-reduction approaches, the main advantage of PNC is that it captures the non-linear structure of the data while retaining size as an explicit coordinate with a clear geometric interpretation, rather than treating size as a nuisance quantity or removing it in preprocessing. The proposed approach also extends to ultra-high-dimensional data by fast PNC. This approximation method is especially useful when the ambient dimension is much larger than the sample size. 


While the methodology in Section~\ref{sec:PNC} is presented in the hypercone setting for clarity, the framework is broadly applicable to more general cone spaces (examples in Section~\ref{sec:conespace}). In this broader view, the radial coordinate represents a generic positive scale parameter, and the base space is a manifold with intrinsic geometry. This generality enables applications beyond classical size-and-shape analysis, including covariance matrices and other structured data with an associated notion of scale (e.g., duration or intensity). Further types of sub-manifold analysis include barycentric subspace analysis \citep{pennec2018barycentric}, principal subsimplex analysis \citep{lee2025principalsubsimplexanalysis}, principal nested Grassmannians \citep{YangVemuri2020_NestedGrassmannians}, and scaled torus PCA \citep{Zoubetal23}, and the extension to the related cone spaces could be useful in applications. 

Several directions for future work remain. First, statistical inference for PNC parameters and scores deserves further development. Second, extending the framework to additional applications, such as longitudinal data, would broaden its use. Third, developing algorithms for more general cone spaces would further expand the scope.

Overall, PNC provides a flexible and interpretable framework for dimension reduction of size--structure data, with a clear geometric interpretation and broad extensions.

\section*{Code availability}
\thispagestyle{plain}
The R code and data are available at \url{https://github.com/YanyanZhan/PNC}.

\section*{Acknowledgments}
\thispagestyle{plain}
The authors thank Professor Charles Laughton (University of Nottingham) for providing the data and the scientific background  
for the sliding clamp multi-protein complex application in Section \ref{sec:moldata}. 


\section*{Disclosure statement}
\thispagestyle{plain}
The authors report there are no competing interests to declare.

\bibliography{extras.bib}

\newpage

\newpage
\appendix

\setcounter{section}{0}
\setcounter{figure}{0}
\setcounter{table}{0}
\setcounter{equation}{0}

\renewcommand{\thefigure}{S\arabic{figure}}
\renewcommand{\thesection}{S\arabic{section}}
\renewcommand{\thesubsection}{S\arabic{section}.\arabic{subsection}}

\begin{center}
{\LARGE \bf Supplementary Material for Principal Nested Cones}
\end{center}

\vspace{1cm}
\section{Proof of Result 4}

The metric tensor on the hypercone is a warped product metric:
$$ds^2 = dr^2 + r^2 \, \sin^2 (\alpha) \,d\Omega^2,$$
where $d\Omega$ is the length element on the unit sphere. 
In order to compute the geodesic, note that 
we can flatten the hypercone 
by transforming the open angle $\alpha$ to 
$\pi/2$. Using the representation of Equation~(4), the first $p-1$ coordinates can be transformed to polar co-ordinates in a $(p-1)$-dimensional Euclidean space. In order to make the mapping
isometric we need to multiply the solid angles 
on the sphere $S^{p-2}$ by $\sin\alpha$, and hence
the flattened space has coordinates: 
\begin{align*}
z_1 &= \mysize \cos(\sin\alpha\; \theta_1) \nonumber \\
z_2 &= \mysize \sin(\sin\alpha\; \theta_1)\cos(\sin\alpha\; \theta_2) \nonumber \\
z_3 &= \mysize \sin(\sin\alpha\; \theta_1)\sin(\sin\alpha\; \theta_2)\cos(\sin\alpha\; \theta_3) \nonumber \\
&\vdots \nonumber \\
z_{p-1} &= \mysize \sin(\sin\alpha\; \theta_1) \; \cdots \; \sin(\sin\alpha\; \theta_{p-2})\cos(\sin\alpha\; \theta_{p-1})
\label{eq:flatcoords}
\end{align*}
The flattened Euclidean space is a hypersector, 
where the angles have reduced range for $\alpha < \pi/2$, 
i.e. $0 \le \theta_i \le \sin \alpha \cdot \pi, i=1,\ldots,d-1, \; \; 0 \le \theta_{d} \le \sin \alpha \cdot 2 \pi$.
Let us write $\Omega_*$ for the part sphere which has angles $\sin\alpha \cdot \theta_i, i=1,\ldots,d$.
The metric tensor on the flattened space is
then 
$$ds^2 = dr^2 + r^2 \; d\Omega_*^2 = dz_1^2 + \ldots dz_{d}^2$$
which is a Euclidean metric. Hence, the geodesic 
distance between two points is given by 
the usual Euclidean distance formula for the Law of Cosines, but with the angular distance on $\Omega_*$, i.e., by Equation~(5). \hfill $\Box$

\newpage

\section{Graphical illustration of Fast-PNC}

\begin{figure}[h!]
    \centering
    \includegraphics[width=.8\linewidth]{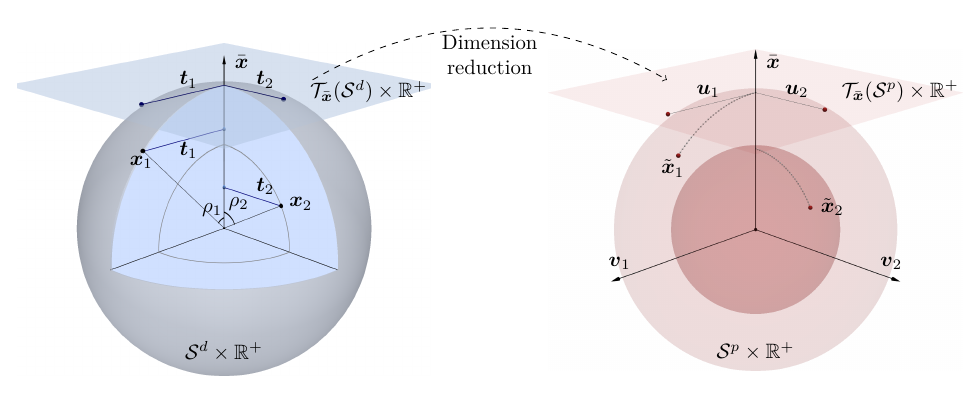}
    \caption{Graphical illustration of PCA-based transformation. The process maps data from 
\(\mathcal{S}^{d} \times \mathbb{R}^{+}\) to its tangent space 
\(\mathcal{T}_{\bar{\boldsymbol{x}}}(\mathcal{S}^{d}) \times \mathbb{R}^{+} \), 
then to \(\mathcal{T}_{\bar{\boldsymbol{x}}}(\mathcal{S}^{p})\times \mathbb{R}^{+}\) through space representation, 
and finally back to \(\mathcal{S}^{p} \times \mathbb{R}^{+}\) through exponential mapping.}
\label{fig:PCA_trans}
\end{figure}

\section{Simulation: spiral data}

\begin{figure}[ht!]
    \centering
    \includegraphics[width=.8\linewidth]{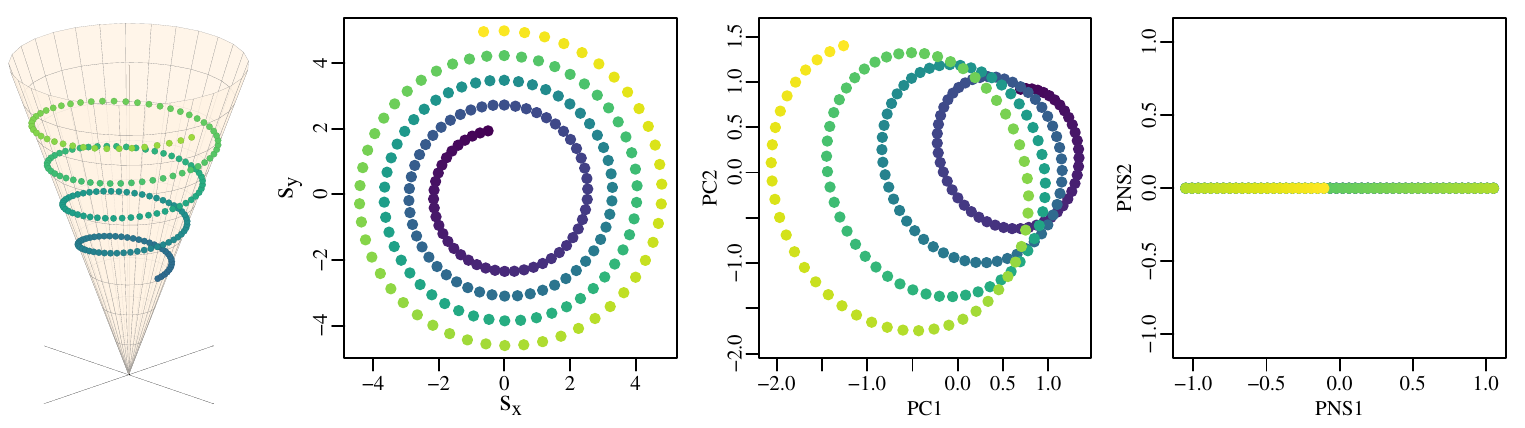}
    \caption{Spiral data on a cone surface with
    \(\alpha = \pi/9\), \(\mysize \in [2,5]\), and \(\theta \in [\pi,8\pi]\) (evenly spaced).
    From left to right: original data, PNC polar scores, PCA scores, and PNS scores. }
    \label{fig:simu2}
\end{figure}

Figure~\ref{fig:simu2} presents a more challenging example in which data lie along a spiral on the cone surface. In this case, PNC (second panel) produces a well-separated and structured representation that reflects the intrinsic geometry of the data. PCA, however, yields an intertwined pattern due to its linear nature (third panel). PNS also fails to recover the spiral structure, as normalization to unit size removes radial variation and collapses the data onto a straight line in the lower-dimensional representation (last panel). As $\alpha$ increases, the cone becomes flatter and the difference between PNC and PCA diminishes gradually, since the underlying manifold becomes closer to Euclidean.

\section{Simulation: back-fitting efficiency}

We compare the representation efficiency and reconstruction performance of PNC, PNS, and PCA using two common metrics: (i) back-fitting distance \(d\), defined as the Euclidean distance between original and reconstructed observations; and (ii) percentage of variance explained \(p\). 
For PCA, $p$ is computed as the proportion of variance explained by the selected components. For PNC and PNS, it is defined as \((\sum_{i=1}^{n} s_{ij}^2)/(\sum_{j=1}^{d}\sum_{i=1}^{n} s_{ij}^2)\).
  

Data are generated uniformly and independently from regions on a cone surface with varying opening angles $\alpha \in \left[ \pi/12, \pi/3 \right]$, and Gaussian noise $\epsilon \sim N(0, \sigma ^2)$ with $ \sigma \in \{0.1, 0.3, 0.5, 1\}$. For each pair $(\sigma, \alpha)$, the experiment is repeated $100$ times to get sample means and $90\%$ confidence intervals.

For three-dimensional data, up to three components can be used for reconstruction. In PCA, components follow the standard PCA framework, whereas in PNC and PNS, they correspond to size, $s_1$ and $s_2$. Although size is not part of the PNS framework, its reconstructions are rescaled using the original size to ensure comparability. Comparisons across the number of components used are not strictly equivalent, due to the differences in component definition. Nonetheless, they provide a useful qualitative assessment.

\begin{figure}[ht!]
	\centering	\includegraphics[width=1\linewidth]{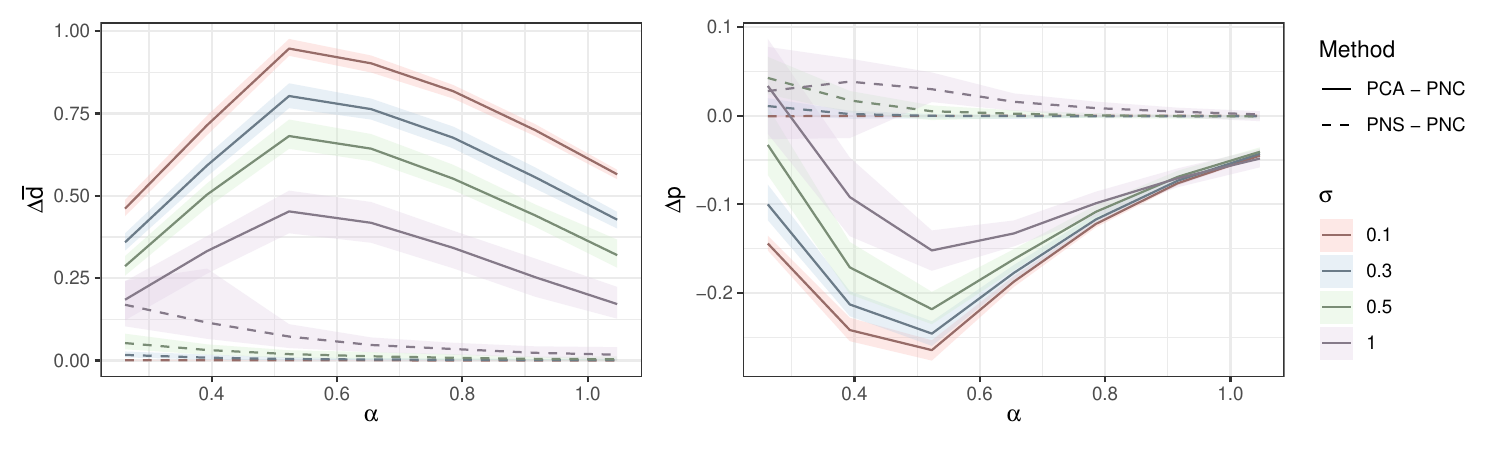}
	\caption{Back-fitting efficiency across opening angles \(\alpha\) and noise levels \(\sigma\). Left: difference in average back-fitting distance. Right: difference in percentage of variance explained. Lines denote means and shaded bands indicate $90\%$ confidence intervals.}
	\label{fig:simu_bf}
\end{figure}

When only one component is used, PCA typically yields smaller reconstruction errors than PNC, since the first PNC component represents size and thus only reconstructs size-specific mean shapes. When all components are included, all methods recover the original data nearly perfectly. We therefore focus on two-component reconstruction comparison, where the differences are most informative. Results are shown in Figure~\ref{fig:simu_bf}.

Across all values of opening angle \(\alpha\) and noise level \(\sigma\), PNC consistently achieves smaller reconstruction error and explains a larger proportion of variance than PCA. 
The performance gap is most pronounced at moderate values of $\alpha$, where the curvature of the manifold is strongest. As $\alpha$ increases and the manifold becomes flatter, the advantage of PNC gradually diminishes, which reflects the fact that Euclidean approximations become more appropriate. Increasing noise levels reduce the overall magnitude of the performance differences but do not change the overall trend.
The performance of PNS remains comparable to that of PNC across all settings, due to the shared underlying curved geometric structure.

\section{Reconstructed molecular deformation}
A video of the reconstructed smooth transition between closed and open conformations is available at  \url{https://github.com/YanyanZhan/PNC/blob/main/output/mol/back-fitting.mp4}

\end{document}